\documentclass[aps,prf]{revtex4-1}
\usepackage{amsmath,amsfonts,bm, color,amssymb}
\usepackage{graphicx,epsfig,overpic,hyperref,ulem}%% OPTIONAL MACRO DEFINITIONS
\input epsf
\UseRawInputEncoding

\newcommand{\eps}{{\varepsilon}}
\newcommand{\sigm}{\sigma}
\newcommand{\out}{{\mathrm{s}}}
\newcommand{\ins}{{\mathrm{d}}}
\newcommand{\ehd}{{\mathrm{ehd}}}
\newcommand{\dep}{{\mathrm{dep}}}
\newcommand{\el}{{\mathrm{el}}}

\newcommand{\visrat}{{\lambda}}
\newcommand{\eq}{{\mathrm{eq}}}

\newcommand{\bs}{\boldsymbol}
\newcommand{\Rr}{\mbox{\it R}}
\newcommand{\Sr}{\mbox{\it S}}
\newcommand{\bE}{{\bf E}}
\newcommand{\bF}{{\bf F}}
\newcommand{\bT}{{\bf T}}
\newcommand{\bU}{{\bf U}}
\newcommand{\bI}{{\bf I}}
\newcommand{\bP}{{\bf P}}
\newcommand{\bR}{{\bf \hat d}}
\newcommand{\bu}{{\bs u}}
\newcommand{\bff}{{\bf f}}
\newcommand{\bt}{{\bf{\hat{t}}}}

\newcommand{\xhat}{{\bf \hat x}}

\newcommand{\zhat}{{\bf \hat z}}

\newcommand{\bx}{{\bf x}}
\newcommand{\by}{{\bf y}}

\newcommand{\bn}{{\bf n}}
\newcommand{\Ca}{\mbox{\it Ca}}

\newcommand{\half}{\frac{1}{2}}

\newcommand{\refeq}[1]{Eq. (\ref{#1})}

\begin{document}

\title{Tandem droplet locomotion in a uniform electric field}
%Cooperative propulsion of  droplets in a uniform electric field}

%\title{Tandem locomotion of droplets in a uniform electric field}
%\title{Droplet propulsion }
%Non-coalescence and cooperative propulsion of  drops in a uniform electric field}
%Non-axisymmetric electrohydrodynamic interaction of drop pairs
\author{Chiara Sorgentone$^{1}$ and Petia M. Vlahovska$^{2}$  }
\affiliation{ $^{1}$ Department of Basic and Applied Sciences for Engineering, Sapienza Universit\`a di Roma, 00161 Rome, Italy\\
$^{2}$ Engineering Sciences and Applied Mathematics, Northwestern University, Evanston, IL 60208, USA. Email: petia.vlahovska@northwestern.edu}

\date{\today}

\begin{abstract}
An isolated charge-neutral droplet in a uniform electric field experiences no net force. However, a droplet pair  can move in response to field-induced dipolar and hydrodynamic interactions. If the droplets are identical, the center of mass of the pair remains fixed. Here, we show that if the droplets have different properties, the pair experiences a net motion due to nonreciprocal %dielectrophoretic and hydrodynamic
 interactions.  
 %We explore the effect of dissimilarity in conductivity and/or permittivity on the tandem droplet locomotion.
%We elucidate that the droplet cooperative propulsion arises from the electrohydrodynamic flow and vanishes in perfect dielectric systems. 
We analyze the three-dimensional droplet trajectories using asymptotic theory, assuming spherical droplets and large separations, and numerical simulations based on a boundary integral method. The dynamics can be quite intricate depending on the initial orientation of the droplets line-of-centers relative to the applied field direction.  
%If initially misaligned with the field, the 
Drops tend to migrate towards a configuration with line-of-centers either parallel or perpendicular to the applied field direction, while either coming into contact or indefinitely separating.  We elucidate the conditions under which these different interaction scenarios take place. Intriguingly, we find 
%In the particular case of droplets aligned 
% we find
 that in some cases droplets can form a stable pair (tandem)  that translates 
%along the field direction while the droplet line-of-centers is 
either parallel or perpendicular to the applied field direction. 

%We study the pair-wise interaction of drops in an applied uniform DC electric field using a combination of numerical simulations based on a boundary integral formulation and an analytical theory assuming small drop deformations.  We study the effect of the difference in the electric properties of the drop fluids, while keeping the drop sizes and viscosities the same, on the drop dynamics. Unlike identical drops, we find that dissimilar drops can form a stable tandem and swim.

%cooperative EHD inerations lead to pripulsion
\end{abstract}

\maketitle

\section{Introduction}
\label{sec:intro}
%While powered by an external electric field, the role of the global uniform field (time-averaged) is to induce a local field gradient in the vicinity of the geometric boundaries of the particle
%The living world abounds with 
%Self-propulsion is the autonomous displacement of nano-, micro- and macroscopic natural and artificial objects, containing their own means of motion.

Electric fields are widely used to steer particles and droplets for applications in directed  assembly \citep{Blaaderen:2013,Harraq:2022}, microfluidics \citep{Link:2006,Hartmann:2022}, ink-jet printing \citep{Basaran2013},  modulation of emulsion microstucture and rheology \citep{Eow:2002,Tao:2016}, and electrosprays \citep{GANANCALVO2018}. 
An important issue in practical applications is the droplet interactions due to electric polarization and electrohydrodynamic flows. In the canonical case of an applied uniform electric field,  
%In the case of identical particles, 
%electric polarization forces due the applied field  favor alignment of the particles line-of-centers with the applied  field direction, and give rise to dielectrophoretic attraction, which in the case of solid particles results in 
the induced dipoles promote particle chaining along the applied field direction
%, which is responsible for the in positive electrorheological effect in suspensions 
\citep{Zukoski:1993,Sheng:2012}.  In addition to the electrostatic interactions, particles may interact electrohydrodynamically due to induced-charge electrophoretic flows in the case of ideally polarizable particles \citep{Squires:2004}  or electric-shear-driven flows about droplets \citep{Melcher-Taylor:1969}.  These flows can be either cooperative or antagonistic to the dipolar interactions \citep{Baygents:1998, DavidS:2008,Park-Saintillan:2010, Chiara:2020}
%Flows driven by the electric field acting on  induced charge near the particle surface as in induced-charge electropho\cite{DavidS:2008,Park-Saintillan:2010} 
%Electroosmotic hydrodynamic flow 
%If the particles are not solid,  a flow driven by electric shear 
% and in the case of droplets promotes coalescence \citep{}.  
%gives rise to electrostatic interactions that favor 
% \cite{ Baygents:1998} was the first to note that the electrohydrodynamic  flow 
% could, however, counteract the dielectrophoretic attraction of particle pairs
  and prevent chaining \citep{Ha-Yang:2000}.  Recently, the three-dimensional interactions of a pair of identical droplets were investigated by means of numerical simulations using the boundary integral method, asymptotic theory for large separations and spherical droplets \citep{Chiara:2019,Chiara:2020, Chiara:2021} and experiments \citep{Kach:2022}. 
The systematic exploration of the effects of fluid properties and the droplet initial configuration revealed intricate relative motions that eventually lead to either droplet coalescence or indefinite repulsion; only if the droplets line-of-centers were initially perpendicular to the applied field direction and the electrohydrodynamic  flow along the droplet surface were equator-to-pole,  the drops motion is eventually arrested and the drops remain at an equilibrium separation. 
%depending on droplet conductivities and permittivities relative to the suspending fluid, and the initial drop pair alignment with the applied electric field.  
%If coalescence is desirable (as in petroleum dehydration) , which if attractive can lead to coalescence but if repulsive 
%and oppositely charged drops have long been assumed to experience an attractive force that favours their coalescenc
%\cite{Baygents:1998} was the first to not 

Asymmetry in terms of droplet size or properties is expected to increase the complexity of the droplet interactions, however the problem has been studied only to a limited extent for small droplet deformations and exploring only effects of size difference \citep{Kach:2022} or only configurations where droplets are aligned with the field \citep{Zabarankin:2020}. Here, we analyze the three-dimensional interactions of dissimilar drops using both theory and simulations. We find novel dynamics such as droplets ``dancing'' , where droplets execute complex trajectories before coming into contact or separating, or ``swimming'', where droplets form a stable pair  that translates in a direction 
either parallel or perpendicular to the applied field.

\section{Problem formulation}

Let us consider 
two  neutrally-buoyant and charge-free drops  with radii $a_i$  and different viscosities $\eta_{\ins,i}$,  conductivities  $\sigm_{\ins,i}$, and permittivities  $\eps_{\ins,i}$, suspended in a fluid with viscosity $\eta_{\out}$,  conductivity $\sigm_{\out}$, and  permittivity  $\eps_{\out}$. 
The mismatch of drop and suspending fluid  properties is characterized by the conductivity, permittivity, and viscosity ratios %is characterized by 
\begin{equation}
\Rr_i=\frac{\sigm_{\ins,i}}{\sigm_{\out}}\,,\quad \Sr_i=\frac{\eps_{\ins,i}}{\eps_{\out}}\,,\quad \lambda_i=\frac{\eta_{\ins,i}}{\eta_\out}\,, \quad i=1,2
\end{equation}
The difference in drop size introduces one more parameter, $\nu=a_2/a_1$.
The distance between the drops' centroids is $d$ and the angle between the drops' line-of-centers with the applied field direction is $\Theta$. The  unit separation
vector between the  drops is defined by the difference between the position vectors of the drops' centers of mass $\bR=(\bx_2^c-\bx^c_1)/d$. The unit vector normal to the drops line-of-centers and orthogonal to $\bR$ is $\bt$.

We adopt the leaky dielectric model, which is widely used to describe the electrohydrodynamics of weakly conducting, viscous fluids \citep{Melcher-Taylor:1969,Saville:1997,Vlahovska:2019}. Fluid motion and electric field are  described by Stokes and  Laplace equations, respectively:
%ssumes creeping flow and charge-free bulk fluids acting as Ohmic conductors.
% {Albeit  an approximation of the actual electrokinetic problem \citep{Saville:1997,Schnitzer-Yariv:2015, Ganan-Calvo:2016, Mori:2018, GANANCALVO2018}, the leaky dielectric model has been successful in modeling many phenomena not only in poorly conducting fluids such as oils, but also  aqueous electrolyte solutions such as in cell-mimicking vesicle systems \citep{Vlahovska-Dimova:2009,Vlahovska:2019}.} 
%The assumption of charge-free fluids decouples the electric and hydrodynamic fields in the bulk. Accordingly,
\begin{equation}
\label{stress_bulk}
\eta \nabla^2\bu-\nabla p=0\,,\quad \nabla\cdot \bE=0\,,
\end{equation}
where   $\bu$ and $p$ are the fluid velocity  and pressure, and $\bE$ is the electric field. Far away from the drops, $\bE^\out\rightarrow \bE^\infty=E_0\zhat$ and $\bu\rightarrow 0$.

At the drop interfaces, normal electric current is continuous, as originally proposed by \cite{Taylor:1966}, $E_n^\out=\Rr E_n^\ins$, where $E_n=\bE\cdot\bn$, 
 and  $\bn$ is the outward pointing normal vector to the drop interface. The surface charge density adjusts to satisfy the current balance, leading to a discontinuity of the displacement field $\eps^\out\left(E_n^\out-\Sr E_n^\ins\right)=q$.

 The electric field acting on the induced surface charge $q$ gives rise to electric shear stress at the interface. The tangential stress balance yields
 %condition for mechanical equilibrium
 \begin{equation}
\label{stress balanceT}
\left(\bI-\bn\bn\right)\cdot \left( \bT^\out- \bT^\ins\right)\cdot\bn+q\bE_t=0 \,, \quad \bx\in \cal{D}\,,
\end{equation}
where  $T_{ij}=-p\delta_{ij}+\eta (\partial_j u_i+\partial_i u_j)$ is the hydrodynamic stress
  and $\delta_{ij}$ is the Kronecker delta function. The electric tractions are calculated from the Maxwell stress tensor $T^\el_{ij}=\eps \left(E_iE_j-E_kE_k\delta_{ij}/2\right)$. $\bE_t=\bE-E_n\bn$ is the  tangential component of the electric field, which is continuous across the interface,  and $\bI$ is the idemfactor. The normal stress balance is
%The hydrodynamic and electric tractions at the interface are discontinuous and balanced by the capillary stress due to curvature along the drop interface
\begin{equation}
\label{stress balance}
\bn \cdot\left( \bT^\out- \bT^\ins\right)\cdot\bn+\half\left(\left(E_n^{\out}\right)^2-\Sr \left(E_n^{\ins}\right)^2-(1-\Sr)E_t^2\right)=\gamma\,\nabla_s\cdot \bn \,, \quad \bx\in \cal{D}\,,
\end{equation}
where  $\gamma$ is the interfacial tension.

Henceforth, all variables are nondimensionalized using the radius of the undeformed drops $a$,  the undisturbed field strength $E_0$, a characteristic applied stress $\tau_c=\eps_\out E_0^2$, and the properties of the  suspending fluid. Accordingly, the time scale is $t_c=\eta_\out/\tau_c$ and  the velocity scale is $u_c=a_1\tau_c/\eta_\out$. 
% The surfactant concentration is normalized by $\Gamma_\eq$ and the interfacial tension - by $\gamma_\eq$.
 The ratio of the magnitude of the electric stresses and surface tension defines the electric capillary number 
 % \begin{equation}
$ \Ca_i=\frac{\eps_\out E_0^2 a_i}{\gamma}\,\,.$
% \end{equation}

\section{Methodology}

Our numerical method and the asymptotic theory for identical drops were presented and validated in \cite{Chiara:2020}. Here we  summarize the extension of  the small-deformation theory and the numerical method to dissimilar drops.\\

{\subsection{Integral representation for the velocity }}
We utilize a Boundary Integral Method (BIM) to solve for the flow and electric fields. Here we derive a boundary integral formulation 
%We modify the boundary integral formulation adopted in \citep{Chiara:2019}
taking into account the fact that the two drops may have different permittivities and conductivities:
\begin{equation} 
\label{eq:BIE01}
\bE^\infty(\bx)+\sum_{j=1}^2 \int_{{\cal{D}}_j} \frac{\hat{\bx}}{4\pi r^3} {\left(\bE^\out(\by)-\bE^i(\by)\right)\cdot\bn(\by)}dS(\by)= \begin{cases} 
\bE^i(\bx)&\mbox{if } \bx $ inside $ {\cal{D}}_i, \\ 
\half \left(\bE^i(\bx)+\bE^\out(\bx)\right) &\mbox{if } \bx \in{\cal{D}}_i, \\ 
\bE^\out(\bx)&\mbox{if } \bx \in S. \\ 
\end{cases}
\end{equation}
where  $\xhat=\bx-\by$ and $r=|\xhat|$. 
The normal and tangential components of the electric field are calculated from the above equation
\begin{equation}
\label{eq:E_n}
\frac{(R_i+1)}{2R_i}E_n(\bx)=\bE^\infty(\bx) \cdot \bn(\bx)+ \sum_{j=1}^2 \frac{\Rr_j-1}{\Rr_j}\bn(\bx)\cdot\int_{{\cal{D}}_j} \frac{\xhat }{4\pi r^3} E_n(\by)dS(\by)\,,
\end{equation}
\begin{equation}
\label{eq:E_t}
\bE_t(\bx)=\frac{\bE^\out(\bx)+\bE^i(\bx)}{2}-\frac{1+\Rr_i}{2\Rr_i}E_n(\bx)\bn(\bx)\,
\end{equation}
for $\bx \in {\cal{D}}_i$. In order to obtain the mean field appearing in eq. (\ref{eq:E_t}) we make use of eq. (\ref{eq:BIE01}) combined with the continuity of normal current across the interface
%eq. (\ref{currencond}):
\begin{equation} 
\label{eq:meanfield}
\half \left(\bE^i(\bx)+\bE^\out(\bx)\right) =
\bE^\infty(\bx)+\sum_{j=1}^2 \int_{{\cal{D}}_j} \frac{\hat{\bx}}{4\pi r^3}{\left(\frac{R_j-1}{R_j}\right)} E_n(\by)dS(\by).
\end{equation}

For the flow field, we have developed the method for fluids of arbitrary viscosity, but for the sake of brevity here we list the equations in the case of equiviscous drops and suspending fluids. The velocity is given by 
\begin{equation}
 \label{eq:main_eq}
 2\bu(\bx)=-\sum_{j=1}^2  \left( \frac{1}{4\pi}\int_{{\cal{D}}_j} \left(\frac{\bff(\by)}{\Ca}-\bff^E(\by)\right)\cdot \left(\frac{\bI}{r}+\frac{\xhat\xhat}{r^3} \right)dS(\by)\right)\,,
 \end{equation}
where $\bff$ and  $\bff^E$ are the interfacial stresses due to surface tension and electric field
\begin{equation}
\bff=\bn\nabla_s \cdot \bn\,,\quad \bff^E=\left(\bE^\out\cdot \bn\right)\bE^\out-\half \left(\bE^\out\cdot\bE^\out\right)\bn-\Sr_i\left(\left(\bE^i\cdot \bn\right)\bE^i-\half  \left(\bE^i\cdot\bE^i\right)\bn\right)\,.
 \end{equation}
{ Drop velocity and centroid are computed from the volume averages}
\begin{equation}
\bU_j=\frac{1}{V}\int_{V_j}\bu dV=\frac{1}{V}\int_{{\cal{D}}_j}\bn\cdot\left(\bu\bx\right) dS\,,\quad \bx^c_j=\frac{1}{V}\int_{V_j}\bx dV=\frac{1}{2V}\int_{{\cal{D}}_j}\bn\left(\bx\cdot \bx\right) dS\,.
\end{equation}

To solve the system of equations
\refeq{eq:E_n}, \refeq{eq:main_eq} we use a Galerkin formulation based on a spherical harmonics representation presented in \cite{Chiara:2019}. All variables (position vector, velocities, electric field) are expanded in spherical harmonics which provides an accurate representation even for relatively low expansion order. In order to deal with the singular and nearly singular integrals that appear in the formulation we evoke specialized quadrature methods able to control the quadrature errors \citep{Chiara:2022},
%A specialized quadrature method for the singular and nearly singular integrals that appear in the formulation and 
and a reparametrization procedure able to ensure a high-quality representation of the drops also under deformation is used to ensure the spectral accuracy of the method \citep{Sorgentone:2018}.

\subsection{Asymptotic theory for at large separations}
\label{theory}

{An isolated, charge-neutral drop in a uniform electric field does not move. The proximity of a boundary \citep{Yariv:2006} or another drop breaks the symmetry and can cause droplet migration.
%pair moves  in response to dipolar and hydrodynamic interactions, 
However if the drops are identical there is no net motion, i.e., their center of mass remains stationary.
%Here we summarize the main results from the asymptotic theory. Details can be found in \cite{Chiara:2020, Kach:2022}.
Here, we apply the asymptotic theory developed in  \citep{Chiara:2020, Kach:2022} to dissimilar drops and show that the asymmetry gives rise to cooperative droplet propulsion. 

%Here we consider the interplay of these }

We first evaluate  the electrostatic interaction of two widely separated spherical drops. In this case,  the drops can be approximated by  point-dipoles. 
The  disturbance field $\bE_1$ of the drop dipole $\bP_1$ induces a dielectrophoretic (DEP) force on the dipole $\bP_2$ located at $\bx^c_2=d\bR$, given by $\bF_2(d)=\left(\bP_2\cdot \nabla \bE_1\right)|_{r=d}$. Likewise, dipole  $\bP_2$ induces a force on dipole 1 that is of equal magnitude and opposite sign $\bF_1=-\bF_2$.
%\begin{equation}
%\bF(d)=\bP_1\bP_2:\nabla \left(\frac{\bI}{r^3}-3\frac{\bx\bx}{r^5}\right)|_{r=d}\,,\quad \bP_1=\bP_2=\frac{\Rr-1}{\Rr+2}\bE^\infty
%\end{equation}
The drop velocity  under the action of this force can be estimated from Stokes law, $\bU_i=\bF_i /\zeta_i$ where $\zeta$ is the friction coefficient $\zeta_i=6\pi (3\lambda_i+2)/(3(\lambda_i+1))$. Thus, 
%The DEP force depends on the angle $\Theta$ between the direction of the external field and the line joining the centers of the two drops
\begin{equation}
\label{DEPF}
%\bU_2^\dep=-\bU_1^\dep\equiv \bU^\dep=2 \frac{\beta_D}{d^4}\left(\frac{3(1+\visrat)}{2+3\visrat}\right)\left[\left(1-3\cos^2\Theta\right)\bR-\sin\left(2\Theta\right)\bt\right]\,,\quad \beta_D= \left(\frac{\Rr_1-1}{\Rr_1+2}\right)\left(\frac{\Rr_2-1}{\Rr_2+2}\right)
\bU_i^\dep=2 \frac{\beta_D}{d^4}\left(\frac{3(1+\visrat_i)}{2+3\visrat_i}\right)\left[\left(1-3\cos^2\Theta\right)\bR-\sin\left(2\Theta\right)\bt\right]\,,\quad \beta_D= \left(\frac{\Rr_1-1}{\Rr_1+2}\right)\left(\frac{\Rr_2-1}{\Rr_2+2}\right)
\end{equation}
If $(\Rr_1-1)(\Rr_2-1)>0$, as in the case of identical droplets, droplets attract if $\Theta<\Theta_c=\arccos\left(\frac{1}{\sqrt{3}}\right)\approx 54.7^o$, e.g., when the drops are lined up with the field, and repel if the line of centers of the two drops is perpendicular to the applied  field. The droplets line-of-centers rotates to align with the applied  field.  However, this situation reverses if $(\Rr_1-1)(\Rr_2-1)<0$: the droplets repel if their  line-of-centers is parallel the applied field direction, and attract if their line-of-centers   is perpendicular to the field. The DEP interaction in this case rotates the droplet line-of-centers away from the applied field direction.
%moves the droplet to a configuration, where 

The electrohydrodynamic (EHD) flow about droplet 1 moves droplet 2 and vice versa the flow about droplet 2 moves droplet 1. The velocities of the droplets  are
\begin{equation}
\label{U2ehd}
    {\bU}^\ehd_{2} = \beta_{T,1} \bU^\ehd\left(d,\visrat_2\right)\,,\quad {\bU}^\ehd_{1} = -\beta_{T,2} \bU^\ehd\left(d,\visrat_1\right)
   \end{equation}
 where
 \begin{equation}   
 \bU^\ehd\left(d,\visrat\right)=   \left(\frac{1}{ d^2}-\frac{2}{d^4}\left(\frac{1+3\visrat}{2+3\visrat}\right)\right)\left(-1+3 \cos^2\Theta\right)\bR-\frac{2}{d^4}\left(\frac{1+3\visrat}{2+3\visrat}\right)\sin(2\Theta)\bt+O(d^{-5})\,.
\end{equation}
and the stresslet magnitude is
\begin{equation}
  \beta_{T,i}=\frac{9}{10}\frac{\Rr_i-\Sr_i}{\left(1+\visrat_i\right)\left(\Rr_i+2\right)^2}\,,\quad i=1,2
\end{equation}
For equiviscous droplets, the relative velocity $\bU_2-\bU_1$ shows that the EHD interaction changes sign (attractive to repulsive or vice versa depending on  $\beta_{T,2}+\beta_{T,1}$) at the same critical angle $\Theta_c$ as the DEP case. However, the EHD interaction also changes sign  at separation $d^2_c=2\left({1+3\visrat}\right)/\left({2+3\visrat}\right)$. $d_c$ ranges from 1 for bubbles ($\visrat=0$) to  $\sqrt{2}$ for very viscous drops ($\visrat\rightarrow\infty$), both corresponding to center-to-center distance smaller than the minimal separation of 2 for spherical drops. Accordingly,  in reality the sign of the EHD interactions does not vary with drop-drop separation.
%The EHD interaction depends on the relative strength of the stresslets. 
 For droplets aligned with the field, both $\beta_T$ negative results in EHD attraction,
 % if $d>d_c$, 
 since the surface flow about each drop is from pole to equator and the fluid is being drawn away from the space between the droplets. Both  $\beta_T$ positive results in repulsion because the surface flow about the droplets is equator to pole and the fluid is being drawn into the space between the droplets, effectively pushing them away.  Dissimilar droplets can either attract or repel depending the relative strength of their stresslets. These scenarios reverse for droplet with line-of-centers perpendicular to the applied field direction.

 \section{Results}
 % so if we put bunch of dissimilar particle we will will create active suspension????
  In this section, we explore the pair-wise droplet dynamics using the analytical theory and numerical simulations.  
 \subsection{Droplet cooperative propulsion}
Dissimilarity creates nonreciprocal interactions which give rise to a net motion of the pair. The ``swimming'' velocity, defined as the velocity of the pair center-of-mass, at leading order is
 \begin{equation}
\bU_s=\frac{1}{2}\left(\bU_2+\bU_1\right)=f(d)\left(-1+3 \cos^2\Theta\right)\bR+g(d) \sin(2\Theta)\bt .
\end{equation}
  The most natural source  of dissimilarity is a difference in droplet size. In this case,  for droplets with same material properties
 \begin{equation}
 \begin{split}
f(d)=&-\frac{\beta_T}{d^2}(\nu^3-1)+ \frac{1}{d^4(2+3\visrat)}\left(\beta_T(1+3\visrat)(\nu^5-1)+3 \beta_D(1+\visrat)(\nu^3-1)\right)\,,\\
g(d)=&\frac{1}{d^4(2+3\visrat)}\left(\beta_T(1+3\visrat)(\nu^5-1)+3 \beta_D(1+\visrat)(\nu^3-1)\right)
\end{split}
\end{equation}
where $\nu$ is the ratio of droplet radii.

Difference in  droplet viscosity also breaks the symmetry and drives self-propulsion. In this case, if all other properties and drop radii are the same,
%If drops are electrically similar but have different viscosities, both the electrohydrodynamic and DEP interactions become nonreciprocal. T
the swimming speed is controlled by the viscosity mismatch of the droplets
\begin{equation}
f(d)=g(d)=\frac{3(\visrat_1-\visrat_2)}{d^4(2+3\visrat_1)(2+3\visrat_2)}\left(-\beta_T+\beta_D\right)\,.
\end{equation}
The swimming direction and speed are controlled by the relative importance of the induced dipole and the EHD stresslet. The EHD flow weakens with increasing conductivity, and for $\Rr\rightarrow\infty$, $-\beta_T+\beta_D\rightarrow 1$. The DEP interaction vanishes at $\Rr=1$, and in this case the swimming is driven by the interaction of the the droplets stresslet flows. 

Here, we focus on droplets with same size and  viscosity but different conductivities and permittivities. In this case, the DEP interactions cancel out
%only the electrohydrodynamic interactions are nonreciprocal 
and the swimming speed is set by the droplet stresslets
% Only in this case, formaton of a stable swimming pair is possible, either for drops line-of-centers aligned or perpendicular to the applied field direction,  and its velocity is
%In the stable-pair mode, the droplets tandem ``swims'' with velocity given by 
 \begin{equation}
\bU_s=\frac{1}{2}\left(\bU_2+\bU_1\right)=\frac{1}{2}\left(\beta_{T,1}-\beta_{T,2}\right)\bU^\ehd\,.
\end{equation}
Hence, the nonreciprocal electrohydrodynamic interaction is the source of the droplet tandem locomotion; the swimming speed vanishes if the droplet stresslets are the same. The direction of motion is determined by the stresslets difference. For example, droplets with $\Rr_1=0.1$, $\Rr_2=100$ and same permittivity ratio ($\Sr_1=\Sr_2=1$) that are initially aligned with the field  translate antiparallel to the field ; swapping the droplets reverses the swimming direction. In this case, droplets settle into a stable separation. In general, however,  the drop pair dynamics is complex because  the center-of-mass motion is superimposed on changes in separation and rotation of the line-of-centers relative to applied field direction.

\subsection{Droplet trajectories}
Here we examine the conditions to form a stable locomoting tandem. 
 According to the theory, the droplet separation and  line-of-center orientation 
  evolve as
 \begin{equation}
 %\dot d=\bU\cdot\bR=\left[\left(\beta_{T,1}+\beta_{T,2}\right)\left(\frac{1}{ d^2}-\frac{2}{d^4}\left(\frac{1+3\visrat}{2+3\visrat}\right)\right)-4 \frac{\beta_D}{d^4}\left(\frac{3(1+\visrat)}{2+3\visrat}\right)\right]\left(-1+3\cos^2\Theta\right)
 \dot d=\bU\cdot\bR=\left[\left(\beta_{T,1}+\beta_{T,2}\right)\frac{1}{ d^2}-\frac{2}{d^4}\Phi\left(\lambda, d\right)\right]\left(-1+3\cos^2\Theta\right)
  \end{equation}
  \begin{equation}
  \label{eqThetatj}
  \dot \Theta=\frac{1}{d}\bU\cdot\bt=-\frac{2}{d^5} \Phi\left(\lambda, d\right)\sin\left(2\Theta\right)
   \end{equation}
   where $\bU=\bU_2-\bU_1$ is the relative velocity and 
  \begin{equation}
  \label{Phif}
  \Phi\left(\lambda, d\right)=\left(\frac{1+3\visrat}{2+3\visrat}\right)\left[\left(\beta_{T,1}+\beta_{T,2}\right)+2\beta_D \left(\frac{3(1+\visrat)}{1+3\visrat}\right)\right]\,.
   \end{equation}
 Examination of this dynamical system shows that there are two equilibrium points: $\Theta_*=0$ and $d_*=d_\eq$, and $\Theta_*=\pi/2$ and $d_*=d_\eq$, where (for viscosity ratio 1)
 \begin{equation}
d^2_\eq=\frac{8}{5}+\frac{32 (\Rr_1-1) (\Rr_1+2) (\Rr_2-1) (\Rr_2+2)}{3 \left(\Rr_1^2
   (\Rr_2-\Sr_2)+\Rr_1 (\Rr_2 (\Rr_2+8)-4 \Sr_2+4)+\Rr_2
   (4-(\Rr_2+4) \Sr_1)-4 (\Sr_1+\Sr_2)\right)}
 \end{equation}
 The equilibrium points are saddles, as seen from the phase plane plotted in Fig. \ref{figPP}.  
  \begin{figure}
 \includegraphics[width=\linewidth]{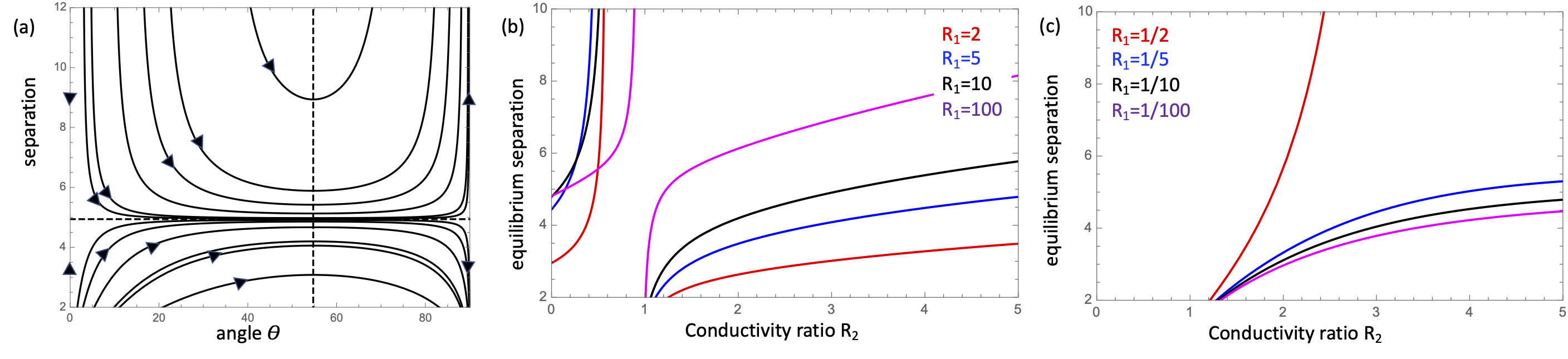}
   \caption{\footnotesize{(a) Phase plane of droplet trajectories for  $\Rr_1=0.1$, $\Rr_2=100$ and $\Sr_1=\Sr_2=1$, corresponding to $\Phi<0$.
   % The $\dot d=0$-nullclines are $d=d_{\eq}$ and $\Theta=\arccos\left(1/\sqrt{2}\right)\approx 57.3^o$. The $\dot\Theta=0$ nullclines are $\Theta=0\,,90^o$. The black dashed lines correspond to the nullclines. Trajectories passing through $d=3,4,6$  and 10 for three different angles $\Theta=5^o,\, 45^o,\, $ and $85^o$ are denoted by black, blue, and red  lines, correspondingly. 
   (b) and (c) Equilibrium separation $d_\eq$ for drops with different conductivity but same permittivity $\Sr_1=\Sr_2=1$. 
   %The dashed line denotes the minimal separation of two droplet radii between the drop centers. 
      %Stable equilibrium separation  for $\Theta=0$ is given by  (c) and the left branch of (b), while for $\Theta=\pi/2$ is the right branch of (b); (c) corresponds to unstable equilibrium. 
      }}
	\label{figPP}
\end{figure}
%Stable pairing occur for drops with line-of-centers at $\Theta_*=0$ or $\Theta_*=\pi/2$, when the droplet properties are such that the DEP is repulsive and stronger than the EHD at short separations and the EHD is attractive at large separations. In such case, drops attract or repel until they reach the equilibrium separation $d_\eq$. If the drops are misaligned but separated exactly by $d_\eq$, their line-of-centers rotates continuously towards either $\Theta_*=0$ , if DEP is weaker than EHD, or $\Theta_*=\pi/2$, in the opposite case, see \refeq{eqThetatj}.
If the droplet line-of-centers is initially aligned with the applied field direction, the droplets attain a steady separation for values of the droplet conductivities corresponding to Fig. \ref{figPP} (c) and the left branch of Fig. \ref{figPP}(b). 
 If the droplet line-of-centers is initially perpendicular to the applied field direction the steady separation is given by the right branch of Fig. \ref{figPP}(b).  In these scenarios, the DEP is repulsive and stronger than the EHD at short separations and the EHD is attractive at large separations. Accordingly, the drops attract or repel until they reach the equilibrium separation $d_\eq$.

Any misalignment of the drops line-of-centers drives the droplets away from the equilibrium configurations towards contact or infinite separation.  The trajectories $d(\Theta)$ are given by
  \begin{equation}
  d^2(\Theta, d_0, \Theta_0)=\frac{f(\Theta, d_0, \Theta_0)}{1+2 b f(\Theta, d_0, \Theta_0)}\,,\quad 
  \end{equation}
 where 
 \[
 f(\Theta, d_0, \Theta_0)=\frac{d_0^2}{1-2 b d_0^2}\left(\frac{\cos\Theta\sin^2 \Theta}{\cos\Theta_0\sin^2 \Theta_0}\right)\,,\quad b=\frac{5(\beta_{T,1}+\beta_{T,2})}{16(\beta_{T,1}+\beta_{T,2}+3\beta_D)}.
 \]
 In the case $\Phi<0$, if the initial separation $d_0>d_\eq$, the droplets  either initially attract but then separate indefinitely if $\Theta_0<\Theta_c$ or monotonically separate if $\Theta_0>\Theta_c$. This scenario is reversed if $d_0<d_\eq$, where droplets ultmately come into contact. 
 %The droplet dynamics is opposite for $\Phi>0$.
 However, if the drops are misaligned but separated exactly by $d_\eq$, i.e., $d=d_0$, their separation remains constant while their line-of-centers rotates continuously towards the equilibrium points, either $\Theta_*=0$ , if $\Phi>0$, or $\Theta_*=\pi/2$, in the opposite case, see \refeq{eqThetatj}.

 \begin{figure}
 \includegraphics[width=\linewidth]{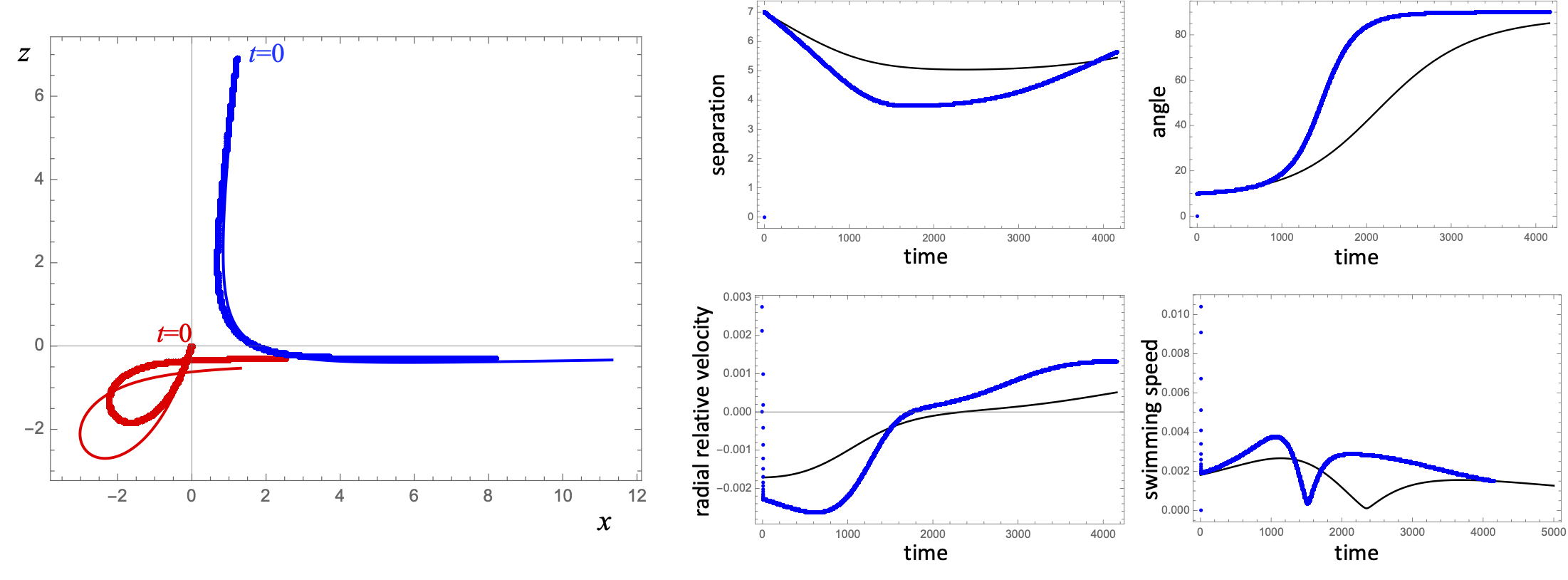}
 \includegraphics[width=\linewidth]{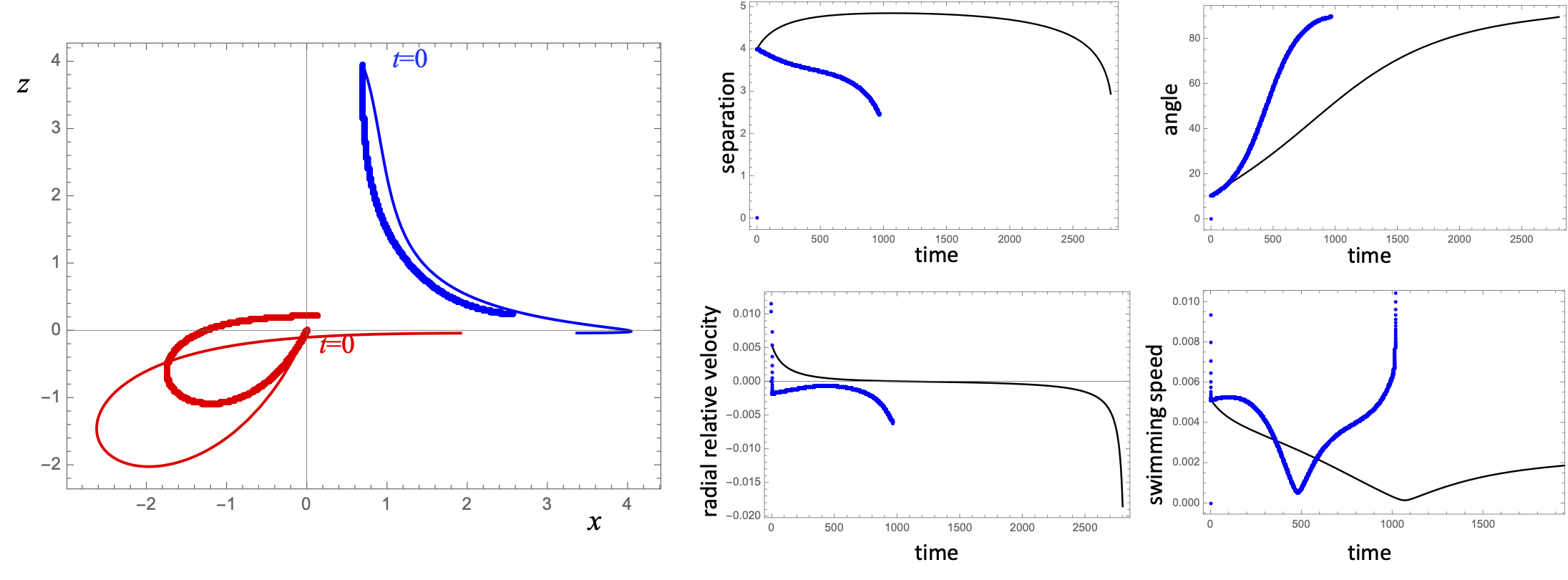}
%          \begin{picture}(0,0)(0,0)
 %   \put(-185,185){(a)}
%\put(20,185){(b)}
%\end{picture}
      \caption{\footnotesize{Droplet-pair dynamics for $\Rr_1=0.1$, $\Rr_2=100$, $\Sr_1=\Sr_2=1$, $\visrat_1=\visrat_2=1$ and initial angle $\Theta_0=10^o$. For this system $d_\eq=4.94$. Lines are computed from the asymptotic theory. Symbols correspond to the numerical simulations.
      Initial separation is $d_0=7$ (top) and $d_0=4$ (bottom). }}
	\label{figT1}
\end{figure}
 Figures \ref{figT1} and \ref{figT3}  illustrate the droplet  dynamics 
 for the cases $\Phi<0$ and $\Phi>0$. If $\Phi<0$ and $\Theta_0=0$ drops form a steady-pair configuration. For $\Theta_0\ne0$, the misaligned droplets migrate towards a configuration where the line-of-centers is nearly perpendicular to the field. In the initial configuration, the induced dipoles are directed in opposite direction, resulting in DEP repulsion.  The stresslets also have opposite sign, however the  EHD flow for the $\Rr=0.1$ droplet is stronger and its pole-to-equator surface flow  results in attraction between the drops. If the initial distance between the drops is greater than the equilibrium separation, $d_0>d_\eq$,  the interaction is initially dominated by the EHD and droplets attract. However, as they get closer the DEP repulsion intensifies and causes them to repel and 
  indefinitely separate, while drop 1 is ``chasing" drop 2, with decreasing swimming speed. If  $\Theta_0<\Theta_c$, the droplets initially attract before starting to repel, see Fig. \ref{figT1}(top); if  $\Theta_0>\Theta_c$ the repulsion is monotonic. 
  In the opposite case, $d_0<d_\eq$, the interactions are reversed: the droplets initially repel and then attract (if $\Theta_0<\Theta_c$), or monotonically attract (if $\Theta_0>\Theta_c$), and eventually come in contact. The droplets relative velocity increases rapidly as they approach each other, see Fig. \ref{figT1}(bottom). The swimming speed varies along the trajectories and it is minimal when the radial velocity is close to zero.  Comparison of the numerical and theoretical results shows that the asymptotic theory qualitatively captures the drop dynamics. The agreement between simulations and theory is better for droplets that are initially farther apart. Thus, given the high computational costs of the simulations, the theory can be used to estimate droplet interactions. 
 Droplet deformation increases the $d_\eq$ above which drops evolve towards separating state. In the considered example, we found by numerical simulations that $d_0=6$ also leads to contact since the deformation causes the drops to get too close and unable to escape the DEP attraction which ultimately leads to contact.

 In the case of droplets pairing in transverse direction, $\Phi>0$, the droplets exhibit the opposite orientational behavior and move to align with the field. If the initial separation is smaller than the equilibrium one, drops come in contact, and otherwise separate indefinitely, while both drops move in opposite direction (''run away" from each other), see Fig. \ref{figT3}. In the latter case, the interaction is extremely weak. The trajectory time is 500000, which is prohibitively expensive to simulate numerically.

  \begin{figure}
\includegraphics[width=\linewidth]{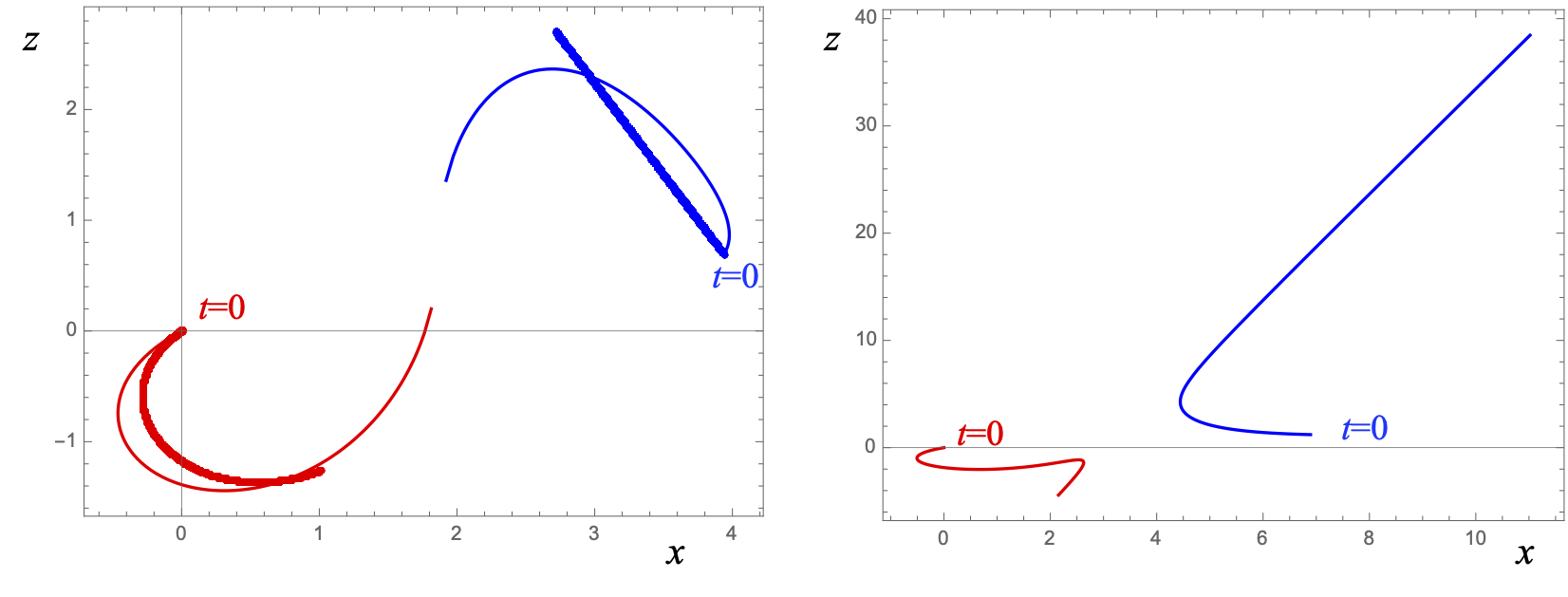}
%          \begin{picture}(0,0)(0,0)
 %   \put(-185,185){(a)}
%\put(20,185){(b)}
%\end{picture}
      \caption{\footnotesize{Droplet-pair dynamics for $\Rr_1=2$, $\Rr_2=100$, $\Sr_1=\Sr_2=1$, $\visrat_1=\visrat_2=1$, and initial angle $\Theta_0=80^o>\Theta_c$. Initial separation (left) $d_0=4$, leading to contact, and (right) $d_0=7$, leading to indefinite separation. Symbols correspond to the numerical simulations. Trajectories  computed from the asymptotic theory are over period of time 2228 (left) and 500000 (right). }	}\label{figT3}
\end{figure}

\section{Conclusions}

We analyze the interactions of dissimilar droplets in a uniform electric field by means of an asymptotic theory, assuming spherical droplets ($\Ca\ll1$) and large separations, and numerical simulations, using a three-dimensional Boundary Integral Method. The  simulations  for $\Ca=0.1$ qualitatively agree with the theory, and thus the theory can be used for a fast estimate of the drop trajectories. Our study focuses on the effect of the mismatch in the electric properties, considering drops with different conductivity and permittivity but same size and viscosity. In this case, the nonreciprocal electrohydrodynamic interactions give rise to a net motion of the drop pair.  The center-of-mass motion is accompanied by changes in drop separation and angle between their line-of-centers to the applied field direction, which gives rise to intricate trajectories. Depending on the droplet stresslets and dipoles in the function $\Phi$, defined by \refeq{Phif}, drops tend to orient their line-of-centers either parallel, if $\Phi>0$, or perpendicular to the applied field direction, if $\Phi<0$.  Initial separation determines if drops will coalesce or indefinitely separate. 
For drops with $\Phi<0$, if $d_0>d_\eq$ and $\Theta_0<\Theta_c$, drops initially attract and then separate indefinitely, while chasing each other. If $d_0<d_\eq$ and $\Theta_0<\Theta_c$, droplets repel and then attract until contact; the interaction is purely attractive and separation decreases  monotonically if $\Theta_0>\Theta_c$. In the particular case of drops aligned with the field and $\Phi<0$, the  drops reach steady separation and ``swim" along the applied field direction, if $\Rr_1<\Rr_2$; direction of motion is reversed if $\Rr_1>\Rr_2$.  If $\Phi>0$ and the drops line-of-centers is perpendicular to the applied field direction the droplet form a tandem swimming transversely to the field. If instead the drops line-of-centers is initially misaligned with the applied field direction, 
with $d_0>d_\eq$ and $\Theta_0>\Theta_c$, drops initially attract, then repel indefinitely while moving in opposite directions of each other. If $d_0<d_\eq$ and $\Theta_0>\Theta_c$, droplets first repel and then attract until contact. In both cases, the  separation changes monotonically if $\Theta_0<\Theta_c$.

Our work represents the first study of the three-dimensional dynamics of electrically dissimilar drops and opens  new directions of exploration of how to manipulate droplets and direct assembly of particles with electric fields.

\section{Acknowledgments}
PV has been supported in part by NSF award  CBET-2126498.  

\section{Declaration of Interests}
The authors report no conflict of interest.

\bibliographystyle{unsrtnat}
%\bibliography{refs2020,refsMY,refs2}

\begin{thebibliography}{27}
\providecommand{\natexlab}[1]{#1}
\providecommand{\url}[1]{\texttt{#1}}
\expandafter\ifx\csname urlstyle\endcsname\relax
  \providecommand{\doi}[1]{doi: #1}\else
  \providecommand{\doi}{doi: \begingroup \urlstyle{rm}\Url}\fi

\bibitem[van Blaaderen et~al.({2013})van Blaaderen, Dijkstra, van Roij, Imhof,
  Kamp, Kwaadgras, Vissers, and Liu]{Blaaderen:2013}
A.~van Blaaderen, M.~Dijkstra, R.~van Roij, A.~Imhof, M.~Kamp, B.~W. Kwaadgras,
  T.~Vissers, and B.~Liu.
\newblock {Manipulating the self assembly of colloids in electric fields}.
\newblock \emph{{Eur. Phys. J -Special Topics}}, {222}\penalty0
  ({11}):\penalty0 {2895--2909}, {NOV} {2013}.
\newblock ISSN 1951-6355.
\newblock \doi{{10.1140/epjst/e2013-02065-0}}.

\bibitem[Harraq et~al.(2022)Harraq, Choudhury, and Bharti]{Harraq:2022}
A.~A. Harraq, B.~D. Choudhury, and B.~Bharti.
\newblock Field-induced assembly and propulsion of colloids.
\newblock \emph{Langmuir}, 38\penalty0 (10):\penalty0 3001--3016, 2022.
\newblock \doi{10.1021/acs.langmuir.1c02581}.
\newblock URL \url{https://doi.org/10.1021/acs.langmuir.1c02581}.
\newblock PMID: 35238204.

\bibitem[Link et~al.(2006)Link, Grasland-Mongrain, Duri, Sarrazin, Cheng,
  Cristobal, Marquez, and Weitz]{Link:2006}
DR~Link, E~Grasland-Mongrain, A~Duri, F~Sarrazin, ZD~Cheng, G~Cristobal,
  M~Marquez, and DA~Weitz.
\newblock Electric control of droplets in microfluidic devices.
\newblock \emph{ANGEWANDTE CHEMIE-INTERNATIONAL EDITION}, 45\penalty0
  (16):\penalty0 2556--2560, 2006.
\newblock ISSN 1433-7851.
\newblock \doi{10.1002/anie.200503540}.

\bibitem[Hartmann et~al.(2022)Hartmann, Schuer, and Hardt]{Hartmann:2022}
Johannes Hartmann, Maximilian~T. Schuer, and Steffen Hardt.
\newblock Manipulation and control of droplets on surfaces in a homogeneous
  electric field.
\newblock \emph{NATURE COMMUNICATIONS}, 13\penalty0 (1), JAN 12 2022.
\newblock \doi{10.1038/s41467-021-27879-0}.

\bibitem[Basaran et~al.({2013})Basaran, Gao, and Bhat]{Basaran2013}
O.~A. Basaran, H.~Gao, and P.~P. Bhat.
\newblock {Nonstandard Inkjets}.
\newblock \emph{{Annu. Review Fluid Mech.}}, {45}:\penalty0 {85--113}, {2013}.
\newblock ISSN {0066-4189}.
\newblock \doi{{10.1146/annurev-fluid-120710-101148}}.

\bibitem[Eow and Ghadiri(2002)]{Eow:2002}
J.~S. Eow and M.~Ghadiri.
\newblock Electrostatic enhancement of coalescence of water droplets in oil: a
  review of the technology.
\newblock \emph{Chem. Eng. Sci.}, 85:\penalty0 357--368, 2002.

\bibitem[Tao et~al.(2016)Tao, Tang, Tawhid-Al-Islam, Du, and Kim]{Tao:2016}
Rongjia Tao, Hong Tang, Kazi Tawhid-Al-Islam, Enpeng Du, and Jeongyoo Kim.
\newblock Electrorheology leads to healthier and tastier chocolate.
\newblock \emph{Proceedings of the National Academy of Sciences}, 113\penalty0
  (27):\penalty0 7399--7402, 2016.
\newblock \doi{10.1073/pnas.1605416113}.
\newblock URL \url{https://www.pnas.org/doi/abs/10.1073/pnas.1605416113}.

\bibitem[Ganan-Calvo et~al.(2018)Ganan-Calvo, Lopez-Herrera, Herrada, Ramos,
  and Montanero]{GANANCALVO2018}
Alfonso~M. Ganan-Calvo, Jose~M. Lopez-Herrera, Miguel~A. Herrada, Antonio
  Ramos, and Jose~M. Montanero.
\newblock Review on the physics of electrospray: From electrokinetics to the
  operating conditions of single and coaxial taylor cone-jets, and ac
  electrospray.
\newblock \emph{Journal of Aerosol Science}, {125}:\penalty0 32--56, 2018.
\newblock ISSN 0021-8502.
\newblock \doi{https://doi.org/10.1016/j.jaerosci.2018.05.002}.
\newblock URL
  \url{http://www.sciencedirect.com/science/article/pii/S0021850217304305}.

\bibitem[Zukoski(1993)]{Zukoski:1993}
C~F Zukoski.
\newblock Material properties and the electrorheological response.
\newblock \emph{Annual Review of Materials Science}, 23\penalty0 (1):\penalty0
  45--78, 1993.
\newblock \doi{10.1146/annurev.ms.23.080193.000401}.
\newblock URL \url{https://doi.org/10.1146/annurev.ms.23.080193.000401}.

\bibitem[Sheng and Wen({2012})]{Sheng:2012}
Ping Sheng and Weijia Wen.
\newblock {Electrorheological Fluids: Mechanisms, Dynamics, and Microfluidics
  Applications}.
\newblock In {Davis, SH and Moin, P}, editor, \emph{{ANNUAL REVIEW OF FLUID
  MECHANICS, VOL 44}}, volume~{44} of \emph{{Annual Review of Fluid
  Mechanics}}, pages {143+}. {2012}.
\newblock ISBN {978-0-8243-0744-8}.
\newblock \doi{{10.1146/annurev-fluid-120710-101024}}.

\bibitem[Squires and Bazant(2004)]{Squires:2004}
T.~M. Squires and M.~Z. Bazant.
\newblock Induced-charge electro-osmosis and electrophoresis.
\newblock \emph{J. Fluid Mech.}, 509:\penalty0 217--252, 2004.

\bibitem[Melcher and Taylor(1969)]{Melcher-Taylor:1969}
J.~R. Melcher and G.~I. Taylor.
\newblock Electrohydrodynamics - a review of role of interfacial shear stress.
\newblock \emph{Annu. Rev. Fluid Mech.}, 1:\penalty0 111--146, 1969.

\bibitem[Baygents et~al.(1998)Baygents, Rivette, and Stone]{Baygents:1998}
J.~C. Baygents, N.~J. Rivette, and H.~A. Stone.
\newblock Electrohydrodynamic deformation and interaction of drop pairs.
\newblock \emph{J. Fluid. Mech.}, 368:\penalty0 359--375, 1998.

\bibitem[Saintillan({2008})]{DavidS:2008}
David Saintillan.
\newblock {Nonlinear interactions in electrophoresis of ideally polarizable
  particles}.
\newblock \emph{{Phys. Fluids}}, {20}\penalty0 ({6}), {JUN} {2008}.
\newblock ISSN {1070-6631}.
\newblock \doi{{10.1063/1.2931689}}.

\bibitem[Park and Saintillan({2010})]{Park-Saintillan:2010}
Jae~Sung Park and David Saintillan.
\newblock {Dipolophoresis in large-scale suspensions of ideally polarizable
  spheres}.
\newblock \emph{{JOURNAL OF FLUID MECHANICS}}, {662}:\penalty0 {66--90}, {NOV
  10} {2010}.
\newblock ISSN {0022-1120}.
\newblock \doi{{10.1017/S0022112010003137}}.

\bibitem[Sorgentone et~al.({2021})Sorgentone, Kach, Khair, Walker, and
  Vlahovska]{Chiara:2020}
C.~Sorgentone, Jeremy~I. Kach, Aditya~S. Khair, Lynn~M. Walker, and Petia~M.
  Vlahovska.
\newblock {Numerical and asymptotic analysis of the three-dimensional
  electrohydrodynamic interactions of drop pairs}.
\newblock \emph{{J. Fluid Mech.}}, {914}:\penalty0 {A24}, {2021}.

\bibitem[Ha and Yang(2000)]{Ha-Yang:2000}
J.-W. Ha and S.-M. Yang.
\newblock Rheological responses of oil-in-oil emulsions in an electric field.
\newblock \emph{J. Rheol.}, 44:\penalty0 235--256, 2000.

\bibitem[Sorgentone et~al.({2019})Sorgentone, Tornberg, and
  Vlahovska]{Chiara:2019}
C.~Sorgentone, A.-K. Tornberg, and Petia~M. Vlahovska.
\newblock {A 3D boundary integral method for the electrohydrodynamics of
  surfactant-covered drops}.
\newblock \emph{{J. Comp. Phys.}}, {389}:\penalty0 { 111--127}, {2019}.

\bibitem[Sorgentone and Vlahovska({2021})]{Chiara:2021}
C.~Sorgentone and Petia~M. Vlahovska.
\newblock {Pairwise interactions of surfactant-covered drops in a uniform
  electric field}.
\newblock \emph{{Phys. Rev. Fluids}}, {5}:\penalty0 {053601}, {2021}.

\bibitem[Kach et~al.(2022)Kach, Walker, and Khair]{Kach:2022}
Jeremy~I. Kach, Lynn~M. Walker, and Aditya~S. Khair.
\newblock Prediction and measurement of leaky dielectric drop interactions.
\newblock \emph{Phys. Rev. Fluids}, 7:\penalty0 013701, Jan 2022.
\newblock \doi{10.1103/PhysRevFluids.7.013701}.
\newblock URL \url{https://link.aps.org/doi/10.1103/PhysRevFluids.7.013701}.

\bibitem[Zabarankin({2020})]{Zabarankin:2020}
Michael Zabarankin.
\newblock {Small deformation theory for two leaky dielectric drops in a uniform
  electric field}.
\newblock \emph{{Proc. Royal Soc. A}}, {476}\penalty0 ({2233}), {JAN 8} {2020}.
\newblock ISSN {1364-5021}.
\newblock \doi{{10.1098/rspa.2019.0517}}.

\bibitem[Saville(1997)]{Saville:1997}
D.~A. Saville.
\newblock Electrohydrodynamics: The {Taylor-Melcher} leaky dielectric model.
\newblock \emph{Annu. Rev.Fluid Mech.}, 29:\penalty0 27--64, 1997.

\bibitem[Vlahovska({2019})]{Vlahovska:2019}
Petia~M. Vlahovska.
\newblock {Electrohydrodynamics of drops and vesicles}.
\newblock \emph{{Annu. Rev. Fluid Mech.}}, {51}:\penalty0 { 305--330}, {2019}.

\bibitem[Taylor(1966)]{Taylor:1966}
G.~I. Taylor.
\newblock Studies in electrohydrodynamics. {I. Circulation} produced in a drop
  by an electric field.
\newblock \emph{Proc. Royal Soc. A}, 291:\penalty0 159--166, 1966.

\bibitem[af~Klinteberg et~al.(2022)af~Klinteberg, Sorgentone, and
  Tornberg]{Chiara:2022}
Ludvig af~Klinteberg, Chiara Sorgentone, and Anna-Karin Tornberg.
\newblock Quadrature error estimates for layer potentials evaluated near curved
  surfaces in three dimensions.
\newblock \emph{Computers \& Mathematics with Applications}, 111:\penalty0
  1--19, 2022.
\newblock ISSN 0898-1221.
\newblock \doi{https://doi.org/10.1016/j.camwa.2022.02.001}.
\newblock URL
  \url{https://www.sciencedirect.com/science/article/pii/S0898122122000517}.

\bibitem[Sorgentone and Tornberg({2018})]{Sorgentone:2018}
C.~Sorgentone and A.-K. Tornberg.
\newblock {A highly accurate boundary integral equation method for
  surfactant-laden drops in 3D}.
\newblock \emph{{J. Comp. Phys.}}, {360}:\penalty0 {167--191}, {MAY 1} {2018}.
\newblock ISSN {0021-9991}.
\newblock \doi{{10.1016/j.jcp.2018.01.033}}.

\bibitem[Yariv(2006)]{Yariv:2006}
Ehud Yariv.
\newblock ``force-free" electrophoresis?
\newblock \emph{Physics of Fluids}, 18\penalty0 (3):\penalty0 031702, 2006.
\newblock \doi{10.1063/1.2185690}.
\newblock URL \url{https://doi.org/10.1063/1.2185690}.

\end{thebibliography}

\end{document}